\documentclass[english,aps,superscriptaddress,twocolumn]{revtex4}
\usepackage[T1]{fontenc}
\usepackage[latin1]{inputenc}
\usepackage{graphicx}
\usepackage{amsmath}
\usepackage{amssymb}

\usepackage{babel}
\makeatother

\begin{document}
\title{Soft x-ray magnetic circular dichroism of LaCoO$_{3}$, La$_{0.7}$Ce$_{0.3}$CoO$_{3}$, and La$_{0.7}$Sr$_{0.3}$CoO$_{3}$ films:
Evidence for Co valency-dependent magnetism and magnetic anisotropy}

\author{M. Merz}
\affiliation{Forschungszentrum Karlsruhe, Institut f\"{u}r Festk\"{o}rperphysik, 76021 Karlsruhe, Germany}
\author{P. Nagel}
\affiliation{Forschungszentrum Karlsruhe, Institut f\"{u}r Festk\"{o}rperphysik, 76021 Karlsruhe, Germany}
\author{C. Pinta}
\affiliation{Forschungszentrum Karlsruhe, Institut f\"{u}r Festk\"{o}rperphysik, 76021 Karlsruhe, Germany}
\affiliation{Physikalisches Institut, Universit\"{a}t Karlsruhe, 76128 Karlsruhe, Germany}
\author{A. Samartsev}
\affiliation{Physikalisches Institut, Universit\"{a}t Karlsruhe, 76128 Karlsruhe, Germany}
\affiliation{Forschungszentrum Karlsruhe, Institut f\"{u}r Festk\"{o}rperphysik, 76021 Karlsruhe, Germany}
\author{H.\@ v.\@ L\"{o}hneysen}
\author{M. Wissinger}
\author{S. Uebe}
\author{A. Assmann}
\affiliation{Forschungszentrum Karlsruhe, Institut f\"{u}r Festk\"{o}rperphysik, 76021 Karlsruhe, Germany}
\affiliation{Physikalisches Institut, Universit\"{a}t Karlsruhe, 76128 Karlsruhe, Germany}
\author{D. Fuchs}
\affiliation{Forschungszentrum Karlsruhe, Institut f\"{u}r Festk\"{o}rperphysik, 76021 Karlsruhe, Germany}

\author{S. Schuppler}
\affiliation{Forschungszentrum Karlsruhe, Institut f\"{u}r Festk\"{o}rperphysik, 76021 Karlsruhe, Germany}
\date{\today}

\begin{abstract}
Epitaxial thin films of undoped LaCoO$_{3}$\@, of electron-doped La$_{0.7}$Ce$_{0.3}$CoO$_{3}$\@, and of hole-doped La$_{0.7}$Sr$_{0.3}$CoO$_{3}$\@ exhibit ferromagnetic order with a transition temperature $T_{\rm{C}}$~$\approx$~85~K, 22~K, and 240~K, respectively. The spin-state structure for these compounds was studied by soft x-ray magnetic circular dichroism at the Co~$L_{2,3}$ and O $K$ edges. It turns out that a higher spin state of the Co$^{3+}$ ions is responsible for the magnetism in LaCoO$_{3}$\@ while for La$_{0.7}$Ce$_{0.3}$CoO$_{3}$ the Co$^{3+}$ ions are in a low-spin state and the ferromagnetism is predominantly determined by Co$^{2+}$\@. For La$_{0.7}$Sr$_{0.3}$CoO$_{3}$\@, on the other hand, the magnetism originates from higher spin states of Co$^{3+}$ and Co$^{4+}$\@. Moreover, a strong magnetic anisotropy  is observed for all systems.
\end{abstract}


\maketitle

Transition-metal oxides with strong electronic correlations have long been known for their exceptional electronic and magnetic phenomena, frequently posing fundamental challenges for our understanding of condensed matter: high-temperature superconductivity  in cuprates or colossal magnetoresistance and complex orbital ordering in manganites are but a few examples. Spearheaded by superconducting Na$_x$CoO$_2 \cdot y$H$_2$O \citep{Takada2003} and by magnetoresistive and orbital ordering effects in $R$BaCo$_2$O$_{5+x}$ \citep{Martin1997}\@, cobaltates have recently emerged as a further center of interest. They are distinguished and even unique in that they offer an additional and possibly tunable aspect {\bf ---} the spin degree of freedom.

Among the many competing interactions occurring on similar energy scales (e.\@g.\@, Hund\textquoteright{}s coupling, crystal field, double exchange, and correlation), the spin degree of freedom is essentially determined by a delicate balance between the crystal-field splitting $\Delta_{\rm CF}$\@, i.\@e.\@, the energetic splitting between $t_{2g}$ and $e_g$ orbitals, and the exchange interaction $J_{\rm ex}$ associated with Hund's rule coupling. Depending on the relative values of $\Delta_{\rm CF}$ and $J_{\rm ex}$, electrons are redistributed between $t_{2g}$ and $e_{g}$ levels, thereby producing in the example of Co$^{3+}$ three possible spin configurations \citep{Korotin1996}: a low-spin state (LS, $t_{2g}^{6}e_{g}^{0}$, $S$=0)\@, a high-spin state (HS, $t{}_{2g}^{4}e_{g}^{2}$, $S$=2), and an intermediate-spin state (IS, $t{}_{2g}^{5}e_{g}^{1}$, $S$=1)\@. In the particular case of bulk LaCoO$_{3}$\@, however, the existence of an IS state is still under debate \citep{Zhuang1998,Noguchi2002,Ropka2003,Haverkort2006,Sundaram2009}\@.

In a previous work \citep{Fuchs2007} it was reported that in contrast to bulk material, epitaxial thin films of LaCoO$_{3}$ become ferromagnetic below the Curie temperature, $T_{\rm{C}}$, of 85~K\@ while polycrystalline thin films grown under similar conditions behave like the bulk material where no indication for ferromagnetism is found down to 5~K\@. It was demonstrated that epitaxial thin films do not show any significant changes for temperatures between 30 K and 450 K, implying that the spin state remains constant and that the tensile strain inhibits a population of the LS state with decreasing temperature \citep{Pinta2008}. Instead, ferromagnetic  order can also be induced in LaCoO$_3$ by partial replacement of La$^{3+}$ by Ce$^{4+}$ or by Sr$^{2+}$\@. A 30~\% substitution by Ce$^{4+}$ (Sr$^{2+}$) leads to a $T_{\rm{C}}$ of 22~K~\citep{Fuchs2005} (240~K~\citep{Fuchs2005b})\@.  This 
doping-dependent discrepancy of $T_{\rm{C}}$ is in contrast to what is found for the manganites where an equivalent amount of electron/hole doping leads for the Mn $t_{2g}^3e_g^{1 \pm x}$ configuration to a comparable change of $T_{\rm{C}}$\@, respectively; for the manganites it was also shown that the ferromagnetic ordering is basically induced by the double-exchange mechanism \citep{Yanagida2009}. Furthermore, doping and epitaxial strain usually modify the Co-O-Co bond angles and the Co-O bond lengths. Since $\Delta_{CF}$\@ is strongly affected by changes in the Co-O bond lengths, the balance between $\Delta_{CF}$\@ and $J_{\rm ex}$ can easily be influenced by epitaxial strain and (electron/hole) doping. Of course, strain and doping effects coexist for doped thin films.

To elucidate the mechanism leading to the magnetic properties in the cobaltates we compare the soft x-ray magnetic circular dichroism (SXMCD) of epitaxially strained LaCoO$_{3}$\@ (LCO), of electron-doped La$_{0.7}$Ce$_{0.3}$CoO$_{3}$ (LCCO), and of hole-doped La$_{0.7}$Sr$_{0.3}$CoO$_{3}$ (LSCO)\@. All film samples were grown on $<$001$>$ oriented 0.1 \% Nb-doped SrTiO$_3$ (Nb:STO) substrates by pulsed laser deposition using stoichiometric sinter targets of the corresponding compound. Film thickness was about 50~nm\@. The growth conditions were the same as those reported in Refs.\ \citep{Fuchs2007,Fuchs2008,Fuchs2005,Fuchs2005b,Pinta2008}\@. The samples were characterized by x-ray diffraction and magnetometry in a SQUID.
Using the setup of the Max-Planck-Institut f\"{u}r Metallforschung (Stuttgart, Germany), SXMCD measurements were performed at 20~K in
total electron yield
and with an applied magnetic field of $\pm 2$~T at the Institut für Festk\"{o}rperphysik beamline WERA at the ANKA synchrotron light source (Karls\-ruhe, Germany)\@. The energy resolution was set to 0.3~eV for the Co $L_{2,3}$ edge and to 0.15~eV for the O $K$ edge. To study $\vec{B} \parallel \vec{n}$ and $\vec{B} \perp \vec{n}$ anisotropies, spectra were taken at normal incidence ($\varphi = \sphericalangle(\vec{n},-\vec{k}) = 0^{\circ}$) and at grazing incidence ($\varphi = 65^{\circ}$)\@, where $\vec{B}$ is the magnetic field, $\vec{n}$ the surface normal of the sample, and $\vec{k}$ the propagation vector of light. $\vec{B} \perp \vec{n}$ values are derived by extrapolating $\varphi$ to $90^{\circ}$ \citep{Stoehr95}. After correction for photon flux variations and for the background, the spectra were normalized at the edge jump, and the finite degree of circular polarization (0.85) was taken into account. Photon energy calibration was ensured by adjusting the Ni $L{}_{3}$ peak position, measured on a NiO single crystal before and after each SXMCD scan, to the established peak position \citep{Reinert1995}\@.

\begin{figure}
\includegraphics[width=0.5\textwidth]{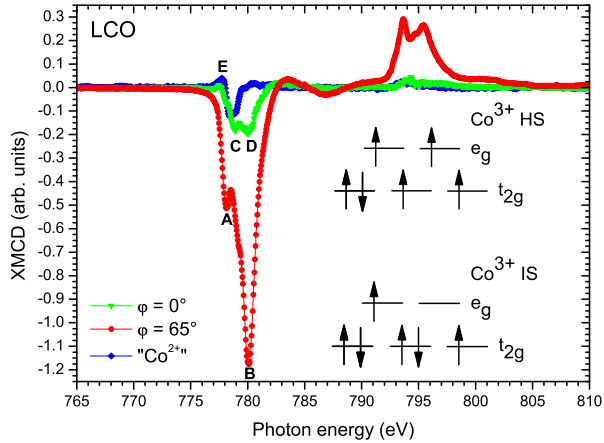}

\caption{\label{fig1}(color online) $\varphi = 0^{\circ}$ and $\varphi = 65^{\circ}$ Co $L_{2,3}$ SXMCD spectra of LCO taken at 20 K. A clear anisotropy between normal and grazing incidence is found. The magnetic ordering is essentially caused by higher spin Co$^{3+}$ states. }

\end{figure}
Fig.\ \ref{fig1} shows the SXMCD spectra of LCO taken at 20~K. A clear anisotropy is found between normal and grazing incidence. Compared with the $\varphi = 0^{\circ}$ spectrum, the $\varphi = 65^{\circ}$ spectrum has a much higher intensity and also the spectral features significantly differ for the two spectra: A distinct double peak is found in the $\varphi = 65^{\circ}$ data with feature A around 778.1~eV and feature B at 780.0~eV, while feature C around 778.9~eV followed by a broader peak at $\approx$ 780.1~eV (feature D) appears for the $\varphi = 0^{\circ}$
\begin{figure}
\includegraphics[width=0.5\textwidth]{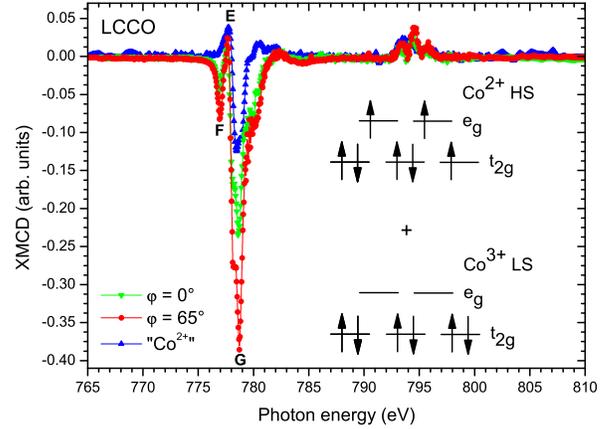}

\caption{\label{fig2}(color online) $\varphi = 0^{\circ}$ and $\varphi = 65^{\circ}$ Co $L_{2,3}$ SXMCD spectra of LCCO taken at 20 K. In contrast to LCO, a reduced anisotropy between $\varphi = 0^{\circ}$ and $\varphi = 65^{\circ}$ spectra is found. The magnetic ordering is predominantly determined by Co$^{2+}$\@.}

\end{figure}
spectrum. Moreover, the SXMCD signal at the $L_2$ edge is almost absent in the $\varphi = 0^{\circ}$ data. This strong anisotropy already illustrates that the easy axis and the magnetic moments of LCO lie predominantly within the film plane.
Additionally, we have included in Fig.\ \ref{fig1} the spectrum of a sample where the only magnetically active sites are Co$^{2+}$ \citep{comment1}. Apparently, Co$^{2+}$ has two characteristic spectral features: An upturn at around 777.7~eV (feature E) and a broad peak centered at 778.5~eV\@. Interestingly, feature E is found in the $\varphi = 0^{\circ}$ spectrum of LCO. Judging from the size of feature E in the $\varphi = 0^{\circ}$ spectrum, a small amount of Co$^{2+}$ is present in LCO due to some residual oxygen deficiency \citep{Pinta2008}. The peak of the ''Co$^{2+}$ spectrum'' at 778.5~eV seems to be partially compensated in the $\varphi = 0^{\circ}$ spectrum by the spectral weight (SW) of feature C. On the other hand, no indications for feature E and, thus, for oxygen deficiency appear in the $\varphi = 65^{\circ}$ data. Taken together, this indicates a negligible contribution of the oxygen-deficient induced Co$^{2+}$ to the total SXMCD SW in the LCO sample, and it is evident from the data shown in Fig.\ \ref{fig1} that the magnetic moment in LCO is essentially determined by higher spin Co$^{3+}$ states. Using the established XMCD sum rules, estimates for the spin and orbital contributions of the magnetic moments can be derived \citep{Stoehr95,Weller95,Chen95}. The values determined in this way for LCO are shown in Table~\ref{tab:table1}. The moments given there and especially the negative value of $\Delta_{\rm orb}=m_{\rm orb}^{\perp}-m_{\rm orb}^{\parallel}$ reflect the strong magnetic anisotropy of the sample with the orbital moment and, as a consequence of the spin-orbit coupling, also the spin mainly residing within the film plane. Employing a single-ion model where $m_{\rm spin} \approx 2 \cdot \langle S_z \rangle$~$\mu_{\rm B}$/Co, the measured spin moment of 1.42~$\mu_{\rm B}$/Co corresponds to $\approx 36$ \% of Co$^{3+}$ in an HS state or to 72 \% of Co$^{3+}$ in an IS configuration. Yet only the first scenario is in agreement with the estimates obtained from multiplet calculations \citep{Pinta2008}\@.
We will continue the discussion of the spin state in the context of the O edge.

In Fig.\ \ref{fig2} the $\varphi = 0^{\circ}$ and $\varphi = 65^{\circ}$ Co $L_{2,3}$ SXMCD spectra of LCCO are depicted. For reference the ``Co$^{2+}$ spectrum'' is again included.
\begin{figure}[t]
\includegraphics[width=0.5\textwidth]{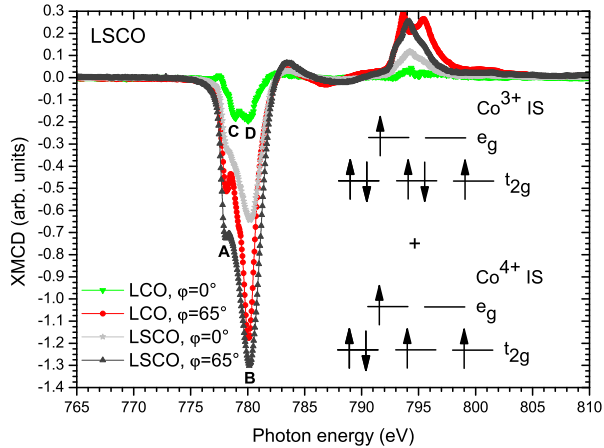}

\caption{\label{fig3}(color online) $\varphi = 0^{\circ}$ and $\varphi = 65^{\circ}$ Co $L_{2,3}$ SXMCD spectra of LSCO taken at 20 K. For comparison the $\varphi = 0^{\circ}$ and $\varphi = 65^{\circ}$ Co $L_{2,3}$ data of LCO are included. The magnetic ordering is caused by higher spin Co$^{3+}$/Co$^{4+}$ states.}

\end{figure}
\begin{table}[b]
\caption{\label{tab:table1} Spin moment $m_{\rm spin}$ and spin density anisotropy $m_{\rm T}^{\parallel}$ and $m_{\rm T}^{\perp}$ together with the orbital moments $m_{\rm orb}^{\parallel}$ and $m_{\rm orb}^{\perp}$\@. $m_{\rm T}^{\perp}$ and $m_{\rm orb}^{\perp}$ are the out-of-plane moments measured at $\varphi = 0^{\circ}$ while $m_{\rm T}^{\parallel}$ and $m_{\rm orb}^{\parallel}$ are the in-plane projected moments calculated from the measured quantities at $\varphi = 0^{\circ}$ and $\varphi = 65^{\circ}$ as discussed in \citep{Stoehr95,Weller95}\@. In addition the averaged orbital moment $m_{\rm orb}^{\rm av}$, the difference $\Delta_{\rm orb}=m_{\rm orb}^{\perp}-m_{\rm orb}^{\parallel}$, and the ratio $m_{\rm orb}^{\rm av}/m_{\rm spin}$ are given.}
\begin{ruledtabular}
\begin{tabular}{lcccc}
                    & LCO        & LCCO    & LSCO     \\ \hline
$m_{\rm spin}$ ($\mu_{\rm B}$/Co)      &  1.422   & 0.856    &  2.500  \\
$m_{\rm T}^{\parallel}$ ($\mu_{\rm B}$/Co)      &  -0.081  & -0.026    & -0.084 \\
$m_{\rm T}^{\perp}$ ($\mu_{\rm B}$/Co)      &  0.171   & 0.065    & 0.168 \\
$m_{\rm orb}^{\parallel}$ ($\mu_{\rm B}$/Co)      &  0.171   & 0.132    & 0.346 \\
$m_{\rm orb}^{\perp}$ ($\mu_{\rm B}$/Co)      &  0.030   & 0.065    & 0.151 \\
$m_{\rm orb}^{\rm av}$ ($\mu_{\rm B}$/Co)      &  0.124   & 0.110    & 0.281 \\
$\Delta_{\rm orb}$ ($\mu_{\rm B}$/Co)      &  -0.142   & -0.067    & -0.196 \\
$m_{\rm orb}^{\rm av}/m_{\rm spin}$      &  0.087  & 0.128    & 0.113 \\
\end{tabular}
\end{ruledtabular}
\end{table}
In contrast to LCO, the anisotropy between the $\varphi = 0^{\circ}$ and $\varphi = 65^{\circ}$ spectra is reduced for LCCO and the spectral shape is very similar. Nevertheless, the still remaining anisotropy reveals that the easy axis and the magnetic moments of LCCO preferably reside in the film plane, too. The spectral shape of the LCCO data clearly differs from that of LCO: a small but conspicuous peak is now observed at 776.9~eV (feature F) followed by the upturn typical for Co$^{2+}$ (feature E described above). A dominant peak is found at $\approx$ 778.7~eV (feature G) with a sharp shoulder around 778.2~eV on the low-energy side of the peak and a broad shoulder on the high-energy side. The features at the Co $L_2$ edge are again quite small. When comparing the main peak at the $L_3$ edge of LCO and LCCO, a clear shift to lower energies by more than 1.3~eV occurs for the latter and the SW in the energy range of the Co$^{3+}$ states is strongly suppressed. This, however, is possible only if the Co$^{3+}$ ion is in an LS state and the magnetic moments of LCCO are dominated by Co$^{2+}$\@. According to the sum rules, the estimates shown in Table~\ref{tab:table1} are deduced for the spin and orbital contributions of the magnetic moments.
Assuming again a single-ion picture, the spin moment of 0.86~$\mu_{\rm B}$/Co is consistent with all 30~\% Co$^{2+}$ in an HS configuration. Correspondingly, the remaining 70~\% Co$^{3+}$ are in an LS state. With this Co$^{2+}$-HS, Co$^{3+}$-LS population the insulating behavior of the compound is also understandable, as the hopping process of the electron from the Co$^{2+}$ to the Co$^{3+}$ site is strongly suppressed by a spin blockade similar to that reported in \citep{Maignan2004}\@. It might be speculated that the Co$^{2+}$ HS state (with its large ionic radius) is stabilized by the tensile strain imposed by the substrate and the (compared to the Co$^{3+}$ HS or IS ion) reduced radius of the Co$^{3+}$ LS state. It seems also worthwile to compare the LCCO data with the ``Co$^{2+}$ spectrum'' shown in Fig.\ \ref{fig2}: while all Co$^{2+}$ states have a characteristic upturn around 777.7~eV (feature E), peak F at 776.9~eV is unique for the Ce-doped samples. This illustrates that the Co$^{2+}$ SXMCD, even though presumably being all HS, is heavily influenced by the chemical surroundings: its shape depends on whether Co$^{2+}$ is induced by oxygen deficiency or by Ce$^{4+}$ doping.

Fig.\ \ref{fig3} shows the $\varphi = 0^{\circ}$ and $\varphi = 65^{\circ}$ Co $L_{2,3}$ SXMCD data of LSCO. For comparison, the corresponding spectra of LCO are included. The two LSCO spectra have a very similar spectral shape: both exhibit a peak at 
780.1~eV (feature B) with a shoulder around 778.0~eV (feature A) on the low-energy side. Hence, the energetic positions of these structures \begin{figure}[t]
\includegraphics[width=0.5\textwidth]{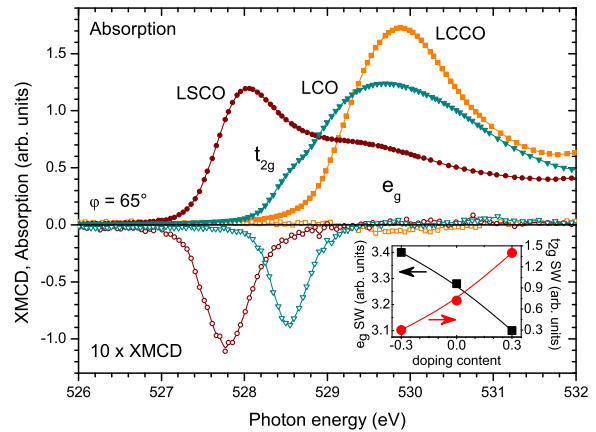}

\caption{\label{fig4}(color online) O $K$ SXMCD and absorption  of LCO, LSCO, and LCCO taken at 20 K. The absorption intensity is strongly doping dependent. A strong XMCD effect is found for LCO and LSCO while it is small for LCCO. Inset: expected SW for $e_g$ and $t_{2g}$ levels (for details, see text). \vspace{-6.663mm}}

\end{figure}
coincide for LSCO and LCO. Yet in contrast to LCO, the $L_2$ edge of LSCO does no longer exhibit a double-peak feature but rather a single peak at 794.1~eV with a shoulder on the low- and high-energy side. Since it is well known that the spectral shape of the $L_2$ edge is very sensitive to the spin state \citep{Chang2009}\@, this finding already points to a different Co$^{3+}$ spin state in LCO and LSCO\@.
Also for LSCO a strong anisotropy with the magnetic moments lying within the film plane is already obvious from the spectra. This is further supported by the estimates for the magnetic moments obtained from the sum rules: again the orbital and spin moment shown in Table \ref{tab:table1} are found to be oriented within the film plane. In a single-ion picture, the spin moment of 2.5~$\mu_{\rm B}$/Co is consistent with 70~\% Co$^{3+}$ in an IS state, 20~\% Co$^{4+}$ in an IS configuration, and 10~\% Co$^{4+}$ in an HS state. Despite the somewhat increased value of the spin moment, the data are compatible with the suggested Co$^{3+}$ IS/Co$^{4+}$ IS picture  \citep{Caciuffo1999,Ravindran2002,Wu2003}\@ and the 2.3 $\mu_{\rm B}$/Co expected for the ideal Co$^{3+}$ IS/Co$^{4+}$ IS configuration: the increased spin moment and, consequently, the 10~\% Co$^{4+}$ found in an HS state seem to be imposed by the tensile strain coexisting with the doping in thin films. It is supposed that the spins couple in LSCO with the charge via the double-exchange mechanism which eventually leads not only to the ferromagnetism but also to the conducting properties of the system.

In Fig. \ref{fig4} the $\varphi = 65^{\circ}$ O $K$ absorption and SXMCD spectra of LCO, LSCO, and LCCO are compared. For LCCO a strong peak centered around 530 eV is found in the absorption data. Using the nomenclature of Ref.\@ \citep{Pinta2008}, this feature corresponds to a large number of unoccupied states in the $e_g$ area. The states in the $t_{2g}$ area (below 529 eV), on the other hand, are almost completely occupied. This result can be qualitatively understood in terms of a very simplified model where charge transfer between Co and O is neglected and just the number of unoccupied states of the $e_g$ and $t_{2g}$ levels is counted: For the 70~\% Co$^{3+}$ LS states both $e_g$ levels are vacant in LCCO and the $t_{2g}$ levels are completely filled; for the 30~\% Co$^{2+}$ HS ions both $e_g$ and one $t_{2g}$ level are half-filled while two $t_{2g}$ levels are filled. Consequently, this configuration leads to the observed strong SW of the $e_g$ states and a quite small intensity for the $t_{2g}$ area. When going to LCO, SW is transferred from the $e_g$ to the $t_{2g}$ range, reflecting a transfer of electrons from $t_{2g}$ to $e_g$ levels. This rearrangement, however, is expected in our simple model for the 36~\% Co$^{3+}$ HS (64~\% Co$^{3+}$ LS) configuration determined for LCO (see inset of Fig.\@ \ref{fig4}) \citep{comment2}\@. For LSCO, the transfer of SW from the $e_g$ to the $t_{2g}$ range continues. Using again our simple model,
the observed transfer of SW is consistent with the 70~\% Co$^{3+}$ IS, 20~\% Co$^{4+}$ IS, and 10~\% Co$^{4+}$ HS configuration discussed above. Thus, the O $K$ absorption data corroborate the different spin states found for LCO, LCCO, and LSCO at the Co $L_{2,3}$ edge.

The O SXMCD spectra in Fig.\@ \ref{fig4} show a strong peak in the $t_{2g}$ area for LSCO and LCO and a very small one in the $e_g$ area for LCCO. For LSCO this feature reflects the transfer of a $t_{2g}$ electron's spin and orbital moment from the Co to the O site concomitant to the double exchange between Co$^{3+}$ and Co$^{4+}$ sites. In the case of LCCO the vanishing SXMCD SW demonstrates that the electrons are very localized due to the spin blockade and do not induce a magnetic moment at the O site by hopping. For strained LCO finally, the peak observed in the $t_{2g}$ area suggests that ferromagnetism is most probably established via superexchange where the spin and orbital moment of $t_{2g}$ electrons is transferred between Co and O. Further investigations will have to determine if this superexchange goes along with orbital ordering.

In conclusion, the present work establishes spectroscopically that due to different spin configurations, ferromagnetism has a different origin in strained LaCoO$_3$, electron-doped La$_{0.7}$Ce$_{0.3}$CoO$_{3}$, and hole-doped La$_{0.7}$Sr$_{0.3}$CoO$_{3}$\@. This gives a natural explanation for the strong discrepancy found for the Curie temperature between the 30 \% electron-doped ($T_{\rm C} \approx 22$~K) and the 30 \% hole-doped ($T_{\rm C} \approx 240$~K) perovskite cobaltate.
\begin{acknowledgments}
Experimental help by B. Scheerer is gratefully acknowledged. We greatly appreciate fruitful discussions with E. Arac, T. Tietze, and E. Goering. We acknowledge the ANKA Ångströmquelle Karlsruhe for the provision of beamtime. Part of this work was supported by the German Science Foundation (DFG) in the framework of the DFG Research Unit 960 ``Quantum Phase Transitions''.
\end{acknowledgments}
\bibliographystyle{apsrev}
\bibliography{PRL}


\end{document}